\documentclass[preprint2]{aastex}
\usepackage{color}

\slugcomment{To appear in Astronomical Journal}

\shorttitle{Kinematics of the Orion Trapezium}
\shortauthors{Olivares et al.}

\begin{document}

\title{Kinematics of the Orion Trapezium based on Diffracto--Astrometry and Historical Data}

\author{J. Olivares\altaffilmark{1},
 L.J. S\'anchez\altaffilmark{1}
A. Ruelas-Mayorga\altaffilmark{1},
        C. Allen\altaffilmark{1},
        R. Costero\altaffilmark{1}, and
        A. Poveda\altaffilmark{1}}

\altaffiltext{1}{Instituto de Astronom\'{\i}a, Universidad
Nacional Aut\'onoma de M\'exico, Apdo. Postal 70--264, 04510,
M\'exico D.F., M\'exico. e-mail: jromero@astro.unam.mx,
leonardo@astro.unam.mx}

\begin{abstract}
Using the novel Diffracto--Astrometry technique we analyze 44
Hubble Space Telescope Wide Field Planetary Camera 2  images of
the Orion Trapezium (OT) taken over a span of 12 years
(1995--2007). We measure the relative positions of the six
brighter OT components (A to F), and supplement these results with
measurements of the relative separations and position angles taken from
the literature, thus extending our analysis time base to $\sim
200$ years. For every pair of components we find the relative rate
of separation as well as the temporal rate of change of their
position angles, which enable us to determine the relative
kinematics of the system. Component E shows a velocity larger that
the OT's escape velocity, thus confirming that it is escaping from
the gravitational pull of this system.
\end{abstract}


\section{INTRODUCTION}
\label{sec:introduction}

With the aim of further checking the dynamical model of
trapezium-type stellar systems proposed by Allen and Poveda
\citep{Poveda1967, Allen1974a}, we decided to apply the novel
Diffracto-Astrometry (DA) technique, developed by
\citet[in preparation]{Sanchez2008,Sanchez2011,Sanchez2013}, to
obtain the relative motions of the most emblematic
trapezium-type system: the Orion Trapezium (OT) = ADS 4186.
In order to do this, we acquired precise measurements of the
relative positions between the six brighter stars (components A to
F) of this system.

In the past, support for this model was found by
\citet{Allen1974b} and \citet{Allen2004}. In both these papers,
studies of the relative motions of trapezium-type systems were
undertaken. The work of these authors along with other
contributions (e.g. \citet{Gualandris2004}) has provided enough
evidence to support this model. Thus, we decided to
perform a study similar to that of \citet{Allen2004} using
historical as well as modern Hubble Space Telescope data.

To obtain precise measurements of the relative positions of the OT
components, we have made use of the high quality images in the HST
Wide Field Planetary Camera 2 (WFPC2) public archive. Due to the
fact that the OT is the trapezium-type system best studied in the
astronomical literature, we also use the historical compilation of
OT data contained in the Washington Double Star Catalog (WDS)
maintained by \citet{Mason2001}.

The majority of the images in the HST--WFPC2 archive that contain
the OT was intended to study the Orion Nebula and the Orion Nebula
Cluster (ONC) and, therefore, have long exposure times which
result, in most cases, in saturated images of the OT
components. This fact, rather than being a disadvantage, works
towards our advantage, because these stellar images show
diffraction spikes which are used by the DA technique to
measure the position of the stars with a precision of $\sim 0.1$
pix \citep{Ruelas2011}. The diffraction spikes are produced by the
supports of the secondary mirrors of both the telescope and the
WFPC2. These effects, together with stellar image saturation,
usually preclude the application of standard astrometric
techniques to these images.

In Section 2 we present a brief description of the DA technique,
Section 3 presents the astrometric data used to obtain the
relative motions of the OT components. Section 4 contains the kinematical results
and, finally, in Section 5 we present our conclusions.

\section{THE DIFFRACTO-ASTROMETRY TECHNIQUE}
\label{sec:DAT}

In the past there have been attempts to perform astrometry as well
as photometry on saturated sources in HST images. For example,
\citet{Golimowski1995} used the diffraction spikes to do
astrometry with a precision of 0.5 pixels;
Gilliland\footnote{Gilliland, R. L. 2007, HST Proposal 11509,
Cycle 16} proposed a way to calibrate the use of the diffraction
spikes to locate the photocenter of a saturated stellar image.
Both \citet{Gilliland1994} and \citet{Maiz2002} have developed
somewhat different methods for obtaining photometry of saturated
sources on HST images. The results appear promising. However, up
to now, a definitive technique for obtaining high precision
astrometric measurements has not been  available.

Diffracto-Astrometry is a technique developed with the intention
of obtaining high precision measurements of relative stellar
positions on saturated stellar images \citep{Sanchez2008}. To
establish a position for a saturated stellar image, DA utilizes in
principle both the diffraction spikes and the diffraction rings on
the images. So far, we have only explored the use of the
diffraction spikes to determine the photocenter of saturated stars
on HST--WFPC2 images of the OT. This technique recovers the
positions of the photocenters taken as the intersection of a pair
of lines fitted to the structure of the diffraction pattern; see
\citet{Sanchez2011} and \citet{Olivares2011} for details.

The application of the DA technique to the measurement of relative
stellar positions on archival HST images requires an algorithm to
ensure minimum systematic errors in the results
\citep[in preparation]{Sanchez2008,Sanchez2011,Sanchez2013}. The
measuring algorithm is designed to detect only those sections of
the spikes that present a large contrast. If at some point the
faint spikes of any component are heavily interfered by the glare
of a brighter component, or if their contrast is too low, the algorithm
simply discards these spikes and the measurement cannot be carried
out. On several images, particularly those with short exposure
times in narrow--band filters, the stellar spikes are not present
for every star or their contrast is too low; thus, it was not
possible to measure every star on every HST image used.

\subsection{Data Corrections}
\label{sec:errors}

There exist a number of systematic errors associated with the
WFPC2 measurements which have been extensively analyzed in the
past. The most important are: geometric distortion
\citep{Kozhurina2003,Anderson2003}, the 34th row error
\citep{Anderson1999}, charge transfer efficiency (CTE)
\citep{Holtzman1995, Stetson1998, Whitmore1999, Dolphin2000}, and
plate scale variations \citep{WFPC2IH,WFPC2DH}.

The OT images used in this analysis were corrected for
geometric distortion and the 34th row error using the formulae given by
the listed above authors. In the case of geometric distortion, there are
published transformation coefficients only for filters F300W,
F555W and F814W. For those images taken with other filters, the
distortion coefficients used were those pertaining to the nearest
filter in wavelength, for which such coefficients exist in the
literature.

It is known that CTE affects astrometry in Advanced Camera for
Surveys Wide Field Channel (ACS/WFC) \citep{Kozhurina2007}.
However, to our knowledge, there is no correction for CTE effects on
WFPC2 astrometry \citep{WFPC2DH}. In addition, according to
\citet[p. 103]{WFPC2IH}, ``Images with high background signal
suffer relatively little charge loss because the background signal
fills the traps and prevents them from robbing the charge packets
during readout.'' Since the OT region is projected over the Orion
Nebula, the background that these images have is relatively high
($\sim 200 \ e^{-}$ for images with exposure times $\geq 10\ s$),
rendering the CTE effect negligible even for the fainter E and F
component spikes.

In order to account for possible changes in plate scale due to the
use of different wavelengths, we have used all the plate scales
reported in Table~5.10 of \citet{WFPC2IH}, and obtained interpolated values for
other filters.

\section{DATA}
\label{sec:data}

In this paper we use OT astrometric data, position angle (PA) and
separation, from two different sources. One is the historical
compilation of the WDS, maintained by \citet{Mason2001}; the
second is data obtained by us applying the DA technique to OT
WFPC2 saturated images. In both these compilations, the PA
values are referred to the North Celestial Pole at their
corresponding epoch. Hence, the measured PAs were precessed to
J2000.0 using the expression given by \citet[p. 276]{Argyle} for objects
located far from the celestial poles, which is the case of the OT.
 The maximum correction applied, to the earliest epoch, amounts to
$\sim 1.0^{\circ}$.

\subsection{Astrometric Data from the Historical Compilation}
\label{sec:WDS} 

Separations and position-angles between OT stellar
pairs were obtained from the WDS. Data in this compilation are
given with no uncertainties, so we selected only those for which
we can assign realistic errors. For this reason, we only used data
for those observers classified as ``Best" and ``Good" in Table 1
of \citet{Allen2004}. As stated by these authors, the error bars in separation
were assigned by C. Worley ``based on a life--time of
experience''. In that study, only the measures for the relative
separations were used, because the historical position angles 
were considered less reliable.

In the present study, the uncertainties for the historical PA observations were
estimated from the dispersions of different measurements by the
same observer, pair of components, and epoch. These measurements
were found to be considerably less reliable than those for the separations.

\subsection{HST--WFPC2 Images}
\label{sec:Images}

The HST public database (Mikulski Archive for Space
Telescopes\footnote{http://archive.stsci.edu/hst/}) was
meticulously inspected in order to find Orion Trapezium images
with the highest astrometric quality. The selection criteria were
the following: (i) We only use images taken with the WFPC2. Since
the WFPC2 is fixed on the axis of the HST, the images taken with it have
a very good optical quality;\, (ii) the OT image must be present
in one of the 4 detectors, to avoid inter-chip errors due to
different reference frames; (iii) the stellar components to be
measured must not lie within the first $\sim 50$ \textit{x} and \textit{y} pixels;
in this region the pyramid mirror makes it impossible to perform
accurate astrometry; (iv) the background of the images must be
$\sim 200 \ e^{-}$ in order to avoid position shifts due to CTE
losses and, (v) the diffraction pattern produced by the supports of
the secondary mirrors (of both telescope and camera), must be
present in at least two images of the OT components.

We used 44 images that fulfilled these criteria. These images were
downloaded in waivered FITS (c0f.fits) format, and were already
calibrated by the standard pipe-line procedure. Their
characteristics are listed in Table~1.

\begin{table}[!ht]
\caption{WFPC2 Images of the Orion Trapezium.}
\resizebox{\columnwidth}{!} {
\small
\begin{tabular}{lcccc}
\hline
Name      & Date & Exps. Time &  Filter & Detector\\
& (UT)    & (secs) & &\\
\hline
          &               &                       &        &  \\
u2id0101t & 3/21/95 5:20   & 180                  & F673N  & 4\\
u2id0104t & 3/21/95 5:39   & 100                  & F631N  & 4\\
u2id0107t & 3/21/95 5:49   & 30                   & F547M  & 4\\
u2id010dt & 3/21/95 7:02   & 180                  & F502N  & 4\\
u2id010ft & 3/21/95 7:12   & 60                   & F656N  & 4\\
          &                &                      &        &  \\
u2id010it & 3/21/95 7:28   & 180                  & F658N  & 4\\
u2id010jt & 3/21/95 8:27   & 180                  & F673N  & 4\\
u2id010nt & 3/21/95 8:52   & 180                  & F631N  & 4\\
u2id010pt & 3/21/95 9:05   & 180                  & F502N  & 4\\
u2id010rt & 3/21/95 10:09  & 180                  & F502N  & 4\\
          &                &                      &        &  \\
u2id010tt* & 3/21/95 10:18  & 30                   & F547M  & 4\\
u2id010vt & 3/21/95 10:23  & 60                   & F656N  & 4\\
u2id010yt & 3/21/95 10:33  & 180                  & F658N  & 4\\
u2id0110t & 3/21/95 11:40  & 180                  & F658N  & 4\\
u2id0111t & 3/21/95 11:49  & 100                  & F673N  & 1\\
          &                &                      &        &  \\
u2id0112t & 3/21/95 11:52  & 100                  & F673N  & 1\\
u2id0113t & 3/21/95 11:55  & 100                  & F673N  & 1\\
u2id0114t & 3/21/95 11:59  & 100                  & F631N  & 1\\
u2id0115t & 3/21/95 12:02  & 100                  & F631N  & 1\\
u2id0118t & 3/21/95 12:12  & 100                  & F502N  & 1\\
          &                &                      &        &  \\
u2id0119t & 3/21/95 12:15  & 100                  & F502N  & 1\\
u2id011at & 3/21/95 12:19  & 30                   & F547M  & 1\\
u2id011et & 3/21/95 13:22  & 60                   & F656N  & 1\\
u2id011gt & 3/21/95 13:29  & 100                  & F658N  & 1\\
u2id011ht & 3/21/95 13:32  & 100                  & F658N  & 1\\
          &                &                      &        &  \\
u3j1030ar* & 4/4/98 5:00    & 160                  & F656N  & 4\\
u4zj0401r & 12/7/98 12:27  & 60                   & F547M  & 1\\
u4zj0402r* & 12/7/98 12:31  & 60                   & F791W  & 1\\
u4zj0405r & 12/7/98 12:49  & 500                  & F547M  & 1\\
u3j1050ar* & 9/20/99 14:30  & 60                   & F547M  & 3\\
          &                &                      &        &  \\
u6c60102r & 3/12/01 22:43  & 30                   & F336W  & 1\\
u6c60108r* & 3/13/01 1:56   & 30                   & F439W  & 1\\
u6eb0102m* & 2/13/02 2:09   & 400                  & F502N  & 4\\
u93k3901m & 4/10/05 17:10  & 80                   & F439W  & 3\\
u93k3902m* & 4/10/05 17:14  & 10                   & F814W  & 3\\
          &                &                      &        &  \\
u93k3904m & 4/10/05 17:28  & 400                  & F336W  & 3\\
u93k3905m & 4/10/05 17:38  & 400                  & F656N  & 3\\
u9w10401m & 11/6/07 13:44  & 40                   & F502N  & 3\\
u9w10403m* & 11/6/07 13:49  & 40                   & F547M  & 3\\
u9w10406m & 11/6/07 13:56  & 40                   & F656N  & 3\\
          &                &                      &        &  \\
u9w10408m & 11/6/07 14:01  & 40                   & F658N  & 3\\
u9w1040am & 11/6/07 14:06  & 40                   & F673N  & 3\\
u9w1040cm & 11/6/07 14:13  & 40                   & F791W  & 3\\
u9w1040em & 11/6/07 14:18  & 40                   & F953N  & 3\\
\hline
\end{tabular}
}
\textsuperscript{*}{\footnotesize Frames used as reference for each epoch listed in Tables 2 and 3.}
\label{t1}
\end{table}

Images are distributed in eight epochs, spanning from 1995 March
21, to 2007 November 6. 56\% of the images were taken in 1995,
16\% in 2007, and the rest during different years of the observing
period (1995--2007). The images have a mean exposure time of $
\sim$120 seconds. 68\%  of them were taken using narrow--band (N)
filters, while 28\% were taken using wide--band (W) and
medium--band (M) filters. 36\% were taken with Detector 4,
27\% with Detector 3, and the rest with Detector 1; no images
taken with Detector 2 were used in this analysis. As an example,
Fig.~1 shows one of the images used in this work.

\begin{figure}[!ht]
\label{imag}
\begin{center}
\includegraphics[width=\columnwidth]{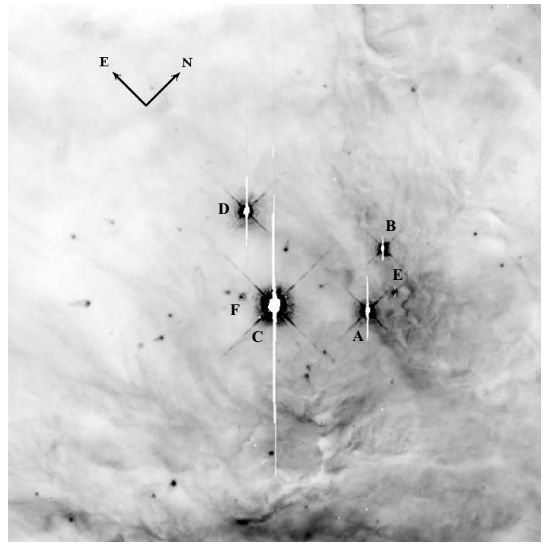}
\caption{Orion Trapezium image (u2id010yt). It was taken using the
WFPC2 Detector 4, in filter F658N, with 180 seconds of exposure
time. The names of the six brighter components have been marked.}
\end{center}
\end{figure}

\subsection{Precision of Diffracto-Astrometry}
\label{sec:Precision}
\citet{Ruelas2011} presented an analysis of the 
 accuracy with which the DA technique is able to recover the
photocenter of a simulated Point Spread Functions (PSFs) of the WFPC2. We 
present here the results of a similar analysis of the DA 
precision on real WFPC2 images. In order to do this, we selected 25 images,
from our data set, taken in one epoch only: 1995 March 21; in this
way we avoid errors associated with individual motions of the
stars. These images were taken with Detectors 1 and 4 (11 and 14
images, respectively) in mid and narrow band filters (3 and 22
images, respectively).

In every one of the WFPC2 images mentioned in the previous
paragraph, the stellar photocenter positions, as well as their
precision, were measured as indicated in Section~2 of
\citet{Ruelas2011}. The precision of all these measurements was
analyzed as a function of detector, filter, spike's effective
Signal-to-Noise (eS/N) and average effective length; 
the last two quantities are defined below.

To determine the eS/N of one single point in the spike, a cut --
21~pix for Detectors 2, 3 and 4 (WF) and 33~pix for Detector 1
(PC) -- perpendicular to the spike at that point is performed. The
point's eS/N is obtained as follows: (i) the length of the cut is
divided in three equal sections; (ii) the average intensity of the two outer
sections is calculated; (iii) the maximum of the central section
is obtained; and (iv) the eS/N is calculated as the ratio of the
maximum value calculated in (iii) to the average value derived in
(ii). The spike's eS/N was computed as the average of the eS/N
values of all the points along the spikes used for establishing a
stellar position.

The average effective length of a spike associated with a
stellar photocenter measurement is defined as the average number
of points along the four spikes used to determine the photocenter
position.

As an example of the precision determination we present, in
Fig.~2, a plot of the distribution of uncertainties as a function of
spike average length. It is clear from this figure that the value
of the precision improves, as expected, with the average length of
the spikes. It seems that the precision may not be improved further
than $\sim0.01$~pix with this technique. However, it was found that the DA
precision is, for all cases, smaller than 0.11 pix.

\begin{figure}[!ht]
\begin{center}
\includegraphics[width=0.5\textwidth]{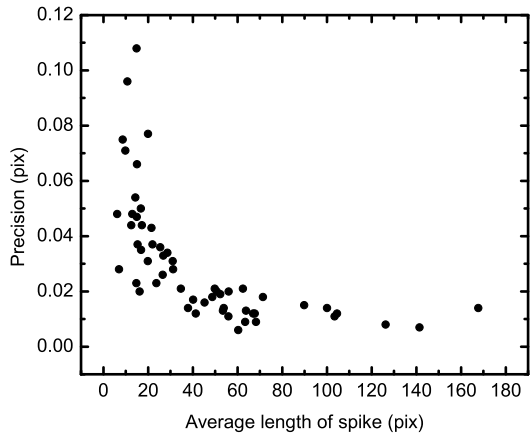}
\caption{Precision of the DA technique as a function of spike
average length. Solid dots represent the measurements for Detector
4 (WF).} \label{prec}
\end{center}
\end{figure}

\subsection{Astrometric Data from the Diffracto-Astrometry Technique}
\label{sec:DA Data}

In each of our 44 images, we measured all present OT
components together with at least 9 (29 at most) unsaturated field
stars. The position  of the saturated stars was measured using the
DA technique, while the position of the unsaturated ones was
estimated using Lorentzian fitting algorithms. Positions were
grouped by epoch and a reference frame was established for each
one of these; we used as a reference frame the image with the
largest number of measured unsaturated positions. In Table~2 we
list the epochs of all the reference frames, together with their
corresponding filter, detector, and plate scale. Using the
{\it{geomap}} task of IRAF\footnote{http://iraf.noao.edu/} and all
unsaturated positions of the epoch reference frame as pivots, we
obtain a transformation between each image and its epoch reference
frame. These transformations allow us to map the positions of the
OT components (A to F) on this epoch pixel-corrected reference
frame. These positions are reported in Table~3.

\begin{table}[!ht]
\caption{Plate-scales for epoch reference frames.}
\begin{center}
\resizebox{\columnwidth}{!} {
\begin{tabular}{ccccc}
Epoch&Filter&Detector&Plate-Scale\textsuperscript{a}&Matched Frames\\
\hline
$1995.21$&F547M&$       4$&$      99.6583$&$25$\textsuperscript{b}\\
$1998.25$&F656N&$       4$&$      99.6923$&$1$\\
$1998.93$&F791W&$       1$&$      45.5733$&$3$\\
$1999.72$&F547M&$       3$&$      99.6147$&$1$\\
$2001.19$&F439W&$       1$&$      45.5468$&$2$\\
$2002.12$&F502N&$       4$&$      99.6631$&$1$\\
$2005.27$&F814W&$       3$&$      99.6529$&$4$\\
$2007.84$&F547M&$       3$&$      99.6147$&$7$\\
\hline
\end{tabular}
}
\end{center}
\textsuperscript{a}{\footnotesize Plate-scales interpolated from
those reported by \citet{WFPC2IH}. Units are mas
pixel\textsuperscript{-1}($\pm 0.005$ for the PC and $\pm 0.01$
for the WF detectors).}\\
\textsuperscript{b} {\footnotesize For this epoch, the 11 PC frames were also matched to WF4 F547M epoch reference frame.}
\label{pltscl}
\end{table}%

\begin{table*}[!ht]
\caption{WFPC2 astrometry of OT components A to F, for different
epochs.}
\begin{center}
\resizebox{2\columnwidth}{!} {
\begin{tabular}{lcccccc}
\hline
Epoch & \multicolumn{6}{c}{Pixel Coordinates (x, y)\textsuperscript{a}} \\
&\multicolumn{2}{c}{A}&\multicolumn{2}{c}{B}&\multicolumn{2}{c}{C}\\
\hline
$1995.21$&$ 417.12 \pm 0.13$&$ 350.18 \pm 0.15$&$ 370.66 \pm 0.11$&$ 426.02 \pm 0.31$&$ 320.88 \pm 0.09$&$ 263.77 \pm 0.17$\\
$1998.25*$&$ 630.03 \pm 0.06$&$ 386.73 \pm 0.06$&$ 583.47 \pm 0.10$&$ 463.08 \pm 0.10$&$ 533.58 \pm 0.01$&$ 300.15 \pm 0.01$\\
$1998.93$&$ 617.05 \pm 0.16$&$ 499.62 \pm 0.14$&$ 515.18 \pm 0.10$&$ 665.80 \pm 0.22$&$ 406.18 \pm 0.19$&$ 311.20 \pm 0.08$\\
$1999.72*$&$ 377.11 \pm 0.01$&$ 321.99 \pm 0.01$&$ 330.25 \pm 0.03$&$ 397.85 \pm 0.03$&$ 280.24 \pm 0.01$&$ 235.65 \pm 0.01$\\
$2001.19$&$               ...           $&$              ...           $&$                 ...         $&$                 ...         $&$ 327.45 \pm 0.13$&$ 222.94 \pm 0.01$\\
$2002.12*$&$ 422.46 \pm 0.08$&$ 426.42 \pm 0.08$&$ 375.82 \pm 0.09$&$ 501.94 \pm 0.09$&$ 326.50 \pm 0.05$&$ 340.06 \pm 0.05$\\
$2005.27$&$ 414.90 \pm 0.11$&$ 380.65 \pm 0.20$&$ 368.48 \pm 0.33$&$ 456.43 \pm 0.18$&$ 318.36 \pm 0.09$&$ 294.16 \pm 0.02$\\
$2007.84$&$ 445.50 \pm 0.12$&$ 532.05 \pm 0.10$&$ 399.15 \pm 0.13$&$ 608.12 \pm 0.08$&$ 349.30 \pm 0.07$&$ 445.51 \pm 0.10$\\
\hline
&\multicolumn{2}{c}{D}&\multicolumn{2}{c}{E}&\multicolumn{2}{c}{F}\\
\hline
$1995.21$&$ 201.95 \pm 0.17$&$ 327.44 \pm 0.12$&$ 424.91 \pm 0.12$&$ 394.80 \pm 0.23$&$ 282.04 \pm 0.26$&$ 240.60 \pm 0.25$\\
$1998.25*$&$ 414.40 \pm 0.05$&$ 364.13 \pm 0.05$&$ 637.92 \pm 0.07$&$ 431.70 \pm 0.07$&$           ...               $&$           ...               $\\
$1998.93$&$ 146.00 \pm 0.10$&$ 450.50 \pm 0.09$&$ 634.38 \pm 0.17$&$ 597.65 \pm 0.20$&$ 321.04 \pm 0.17$&$ 260.97 \pm 0.08$\\
$1999.72*$&$ 161.34 \pm 0.01$&$ 299.43 \pm 0.01$&$ 385.59 \pm 0.05$&$ 366.91 \pm 0.05$&$           ...               $&$           ...               $\\
$2001.19$&$ 67.35 \pm 0.06  $&$ 362.21 \pm 0.08$&$           ...               $&$           ...               $&$ 241.64 \pm 0.08$&$ 171.85 \pm 0.14$\\
$2002.12*$&$ 207.97 \pm 0.11$&$ 403.78 \pm 0.11$&$           ...               $&$           ...               $&$           ...               $&$           ...               $\\
$2005.27$&$ 199.28 \pm 0.41$&$ 358.01 \pm 0.24$&$ 422.69 \pm 0.10$&$ 425.18 \pm 0.12$&$ 278.86 \pm 0.23$&$ 271.07 \pm 0.04$\\
$2007.84$&$ 230.63 \pm 0.16$&$ 509.41 \pm 0.13$&$ 453.15 \pm 0.05$&$ 576.62 \pm 0.04$&$           ...               $&$           ...               $\\
\hline
\end{tabular}}
\end{center}
\textsuperscript{a} Coordinates (x and y) in the system of epoch reference frame, corrected for geometric distortions. Increasing y is to the North, and increasing x is to the
West. Uncertainties are given as the standard deviation, except for those cases marked by an asterisk, in which they correspond to the DA technique precision. 
\label{default}
\end{table*}

Using the positions of the photocenters reported in Table~3,
together with their corresponding plate-scales in Table~2, we
were able to measure, for every pair of components, the separation
(in arcseconds) and the PA (in degrees, measured by convention,
from North to East).

The North direction was established using the group keyword
ORIENTAT  given in the header of each WFPC2 image. This keyword
records the angle, in degrees,  between the North and the columns
of the detector; it is measured from North to East.
Determination of the North direction is limited by the HST guide
stars positions, the mode in which these stars were acquired by
the Fine Guidance Sensors (FGS), and the transformation between
FGS and WFPC2. In all of our images, the FGS's acquisition mode
was FINE; this mode renders a typical rms pointing precision
better than 2 to 5 mas \citep{MDH}. These figures, transformed to
the WFPC2, provide an absolute astrometry accuracy between 0.5 and 1
arcsec \citep{WFPC2DH}.

The entire list of measurements, both from WDS and DA, can be
obtained from the authors by request.

A combination of OT--astrometric data from the WDS and
DA-measurements of 44 OT HST-images allows us to present in the
next section the internal motions of the OT.

\section{KINEMATICS OF THE ORION TRAPEZIUM}
\label{sec:Kinematics}
\subsection{Astrometric Results}

To determine the internal kinematics of the OT components, we
performed linear least-squares (LSQ) fits to the
separation and the PA, both for WDS as well as DA data, as
a function of time.

As examples of the relative motions of the OT components, we
present Figs.~3 to 6 in which we plot the values for the separation and the PA of the component pairs AC,
AE, CB and CD, as a function of time. The straight line resulted from linear LSQ--fits
to the points. 

Since the precision of the PA historical measurements is low, as stated in Section 3.1, and the time base of the DA measures is relatively short, the rates of change derived from the fit to the PA data are significantly less reliable than those obtained for the fits to the separation data.

\begin{figure*}[p]
\begin{center}
\includegraphics[width=\textwidth]{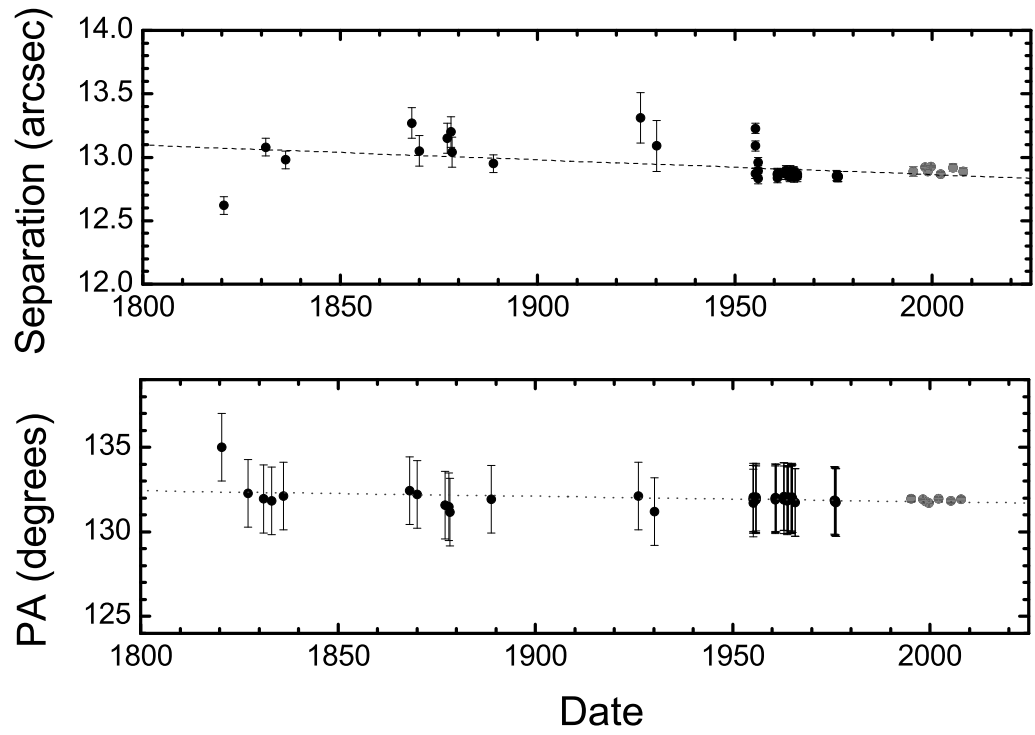}
\caption{Separation and precessed PA of component C relative to A,
as a function of time. Black circles correspond to WDS data with
its associated error bars. Gray circles denote DA data with its
associated error bars (smaller than symbol). The lines show
the best LSQ--fit to the data.}
\end{center}
\label{figac}
\end{figure*}

\begin{figure*}[p]
\begin{center}
\includegraphics[width=\textwidth]{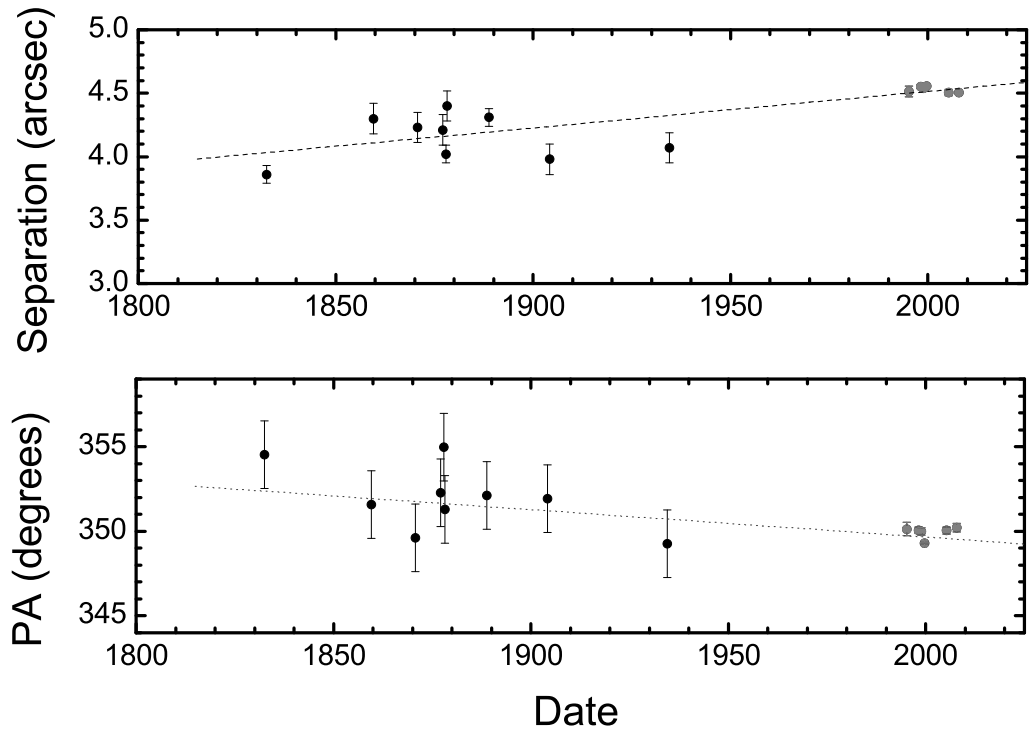}
\caption{Separation and precessed PA of component E with respect
to component A as a function of time. Symbols and lines as in
Fig.~3. Complementing this information with radial velocity measurements
from \citet{Costero2008}, we confirm that component E is escaping from the
Trapezium.}
\end{center}
\label{figae}
\end{figure*}

\begin{figure*}[p]
\begin{center}
\includegraphics[width=\textwidth]{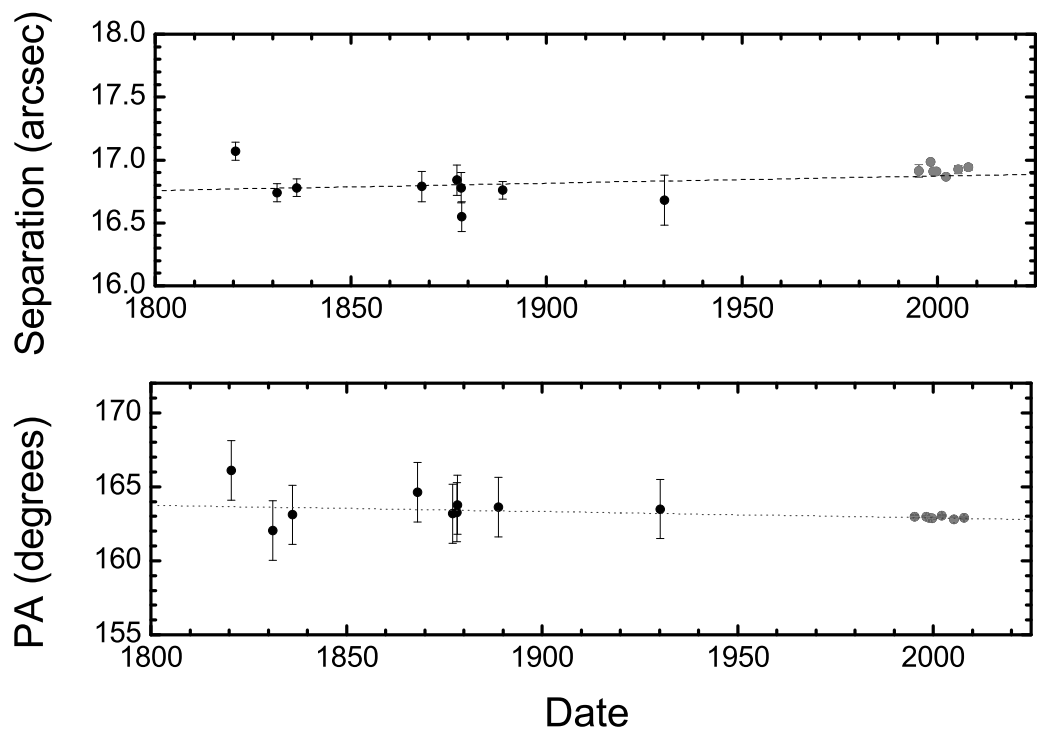}
\caption{Separation and precessed PA of component B with respect
to component C as a function of time. Symbols and lines as in
Fig.~3.}
\end{center}
\label{fibcb}
\end{figure*}

\begin{figure*}[p]
\begin{center}
\includegraphics[width=\textwidth]{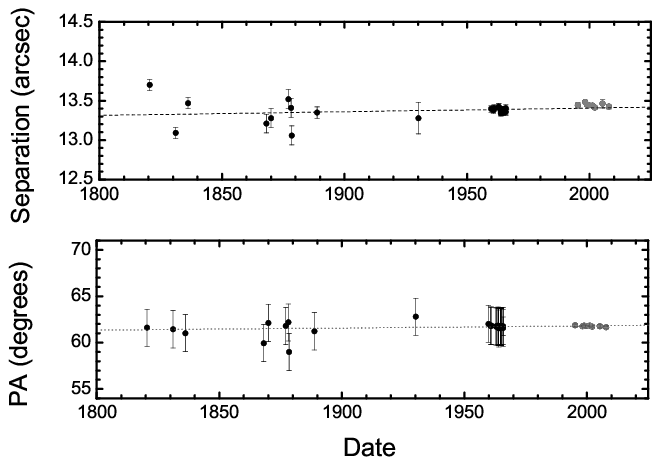}
\caption{Separation and precessed PA of component D with respect
to component C as a function of time. Symbols and lines as in
Fig.~3.}
\end{center}
\label{figcd}
\end{figure*}

Tests for membership of OT components A to E to the Trapezium were
already performed by \citet{Allen2004}. However, in that study it was not possible
 to establish the physical membership of component F to the
OT, due to the fact that component C shows widely discrepant
values for its proper motion \citep{Jeffers1963, Mason2001} 
and hence the proper-motion test for pair CF could not be carried out. Given
the large transverse velocity of F relative to C derived from our measures for the PA and separation velocities ($11.2 \pm 0.8$ km s\textsuperscript{-1} and $2.7 \pm 0.9 $ km s\textsuperscript{-1}, respectively) we think it is a
foreground star. It is probably related to the 
stellar population located in front of the ONC, as proposed by \citet{Alves2012}. 
In view of this fact, we decided to disregard the
kinematical information of this component.

The relative motions of components A to E of the OT are summarized in Table~4.
In this Table, the first column indicates the pair of OT components, the second and third the rate of change in PA and separation, respectively. The figures in the last two columns were transformed to km s\textsuperscript{-1} adopting a
distance to the Orion Nebula of $414\pm7$ pc \citep{Menten2007}.
The precision of these relative velocity
determinations was found to be $\sim1.0$ km s\textsuperscript{-1}; this value corresponds to the
average of uncertainties in Table~4.

\begin{table}[h!]
\begin{center}
\caption{Temporal rate of change of PA and Separation.}
\resizebox{\columnwidth}{!} {
\begin{tabular}{crr}
\hline
Pair           & PA Velocity                              & Separation Velocity       \\
        &(km\ s\textsuperscript{-1} )        &  (km\ s\textsuperscript{-1} )        \\
\hline
AB  &$   -0.7  \pm   0.8  $&$    4.1  \pm   0.5  $\\
AC  &$   -1.4  \pm   0.5  $&$   -2.3  \pm   0.7  $\\
AD  &$   -1.5  \pm   0.4  $&$    1.6  \pm   0.4  $\\
AE  &$   -2.9  \pm   0.9  $&$    5.5  \pm   1.2  $\\
BC  &$   -3.4  \pm   1.7  $&$    1.4  \pm   0.9  $\\
BD  &$   -0.5  \pm   0.7  $&$    2.8  \pm   0.4  $\\
BE  &$    0.5  \pm   0.8  $&$   -1.4  \pm   0.9  $\\
CD  &$    1.5  \pm   0.7  $&$    0.9  \pm   0.6  $\\
CE  &$   -0.4  \pm   1.0  $&$   -5.7  \pm   0.8  $\\
DE  &$    1.9  \pm   1.0  $&$    5.0  \pm   1.2  $\\

\hline
\end{tabular}
}
\end{center}
\label{tk}
\end{table}

Finally, we note that the results presented in this section are
fairly consistent with those published by \citet{Allen2004}. These
authors determine the magnitude of the relative motions
between pairs AB, AC and AE as +0.18, -0.11, and +0.36 seconds of
arc per 100 years, respectively, disregarding the changes in PA.
A transformation of these motions
to transverse velocities results in +3.6, -2.2, and +7.1 km
s$^{-1}$, respectively, again using $ 414 \pm 7$ pc for
the distance to the Orion Nebula \citep{Menten2007}.

\subsection{Heliocentric Radial Velocities and Multiplicity}
\label{sec:RVs}

$\theta^1$ Ori A is a spectroscopic and eclipsing binary (V1016
Ori). Its systemic radial velocity has been obtained by
\citet{Vitrichenko1998} and \citet{Stickland2000}. Based on
essentially the same previously published data and a few
additional data points of their own, both groups obtain the value
given in Table~5. The star is, in addition, an interferometric
binary under close scrutiny (see \citet{Preibisch1999,Close2012}).

Itself a spectroscopic and eclipsing binary, the orbital
parameters of $\theta^1$ Ori B (BM Ori) have been obtained by
several authors. \citet{Vitrichenko2004} notice that the values of
the systemic velocity published for this system differ
significantly between authors and, hence, between epochs. To
reconcile previous results with their own, these  authors
postulate a third component in a 3.57 year, highly eccentric
orbit. In Table 5 we have adopted the systemic velocity of the
putative triple system, as given by \citet{Vitrichenko2004}. This
system is part of a ``mini-cluster"; see \citet{Close2012}.

$\theta^1$ Ori C, the brightest member of the Trapezium, is an
interferometric binary whose primary component is probably a
spectroscopic binary system \citep{Vitrichenko2002a}. Complicated
by line-profile periodical variability associated with the oblique
magnetic rotator character of the brightest component, no reliable
systemic radial velocity is available for this binary. We have
adopted 23.6 km s\textsuperscript{-1}, a value obtained by
\citet{Kraus2009} by adjusting published radial velocity values
for the primary component to that derived from the interferometric
binary orbit. No error was associated to this barycentric velocity
in that paper, but its value is nearly equal to the average of
those obtained recently by \citet{Lehmann2010} and
\citet{Vitrichenko2012} for the close binary ($P \simeq 61 \ d$).
A conservative guess of the error associated with it is $\pm \ 2$
km s\textsuperscript{-1}.

It is frequently assumed that $\theta^1$ Ori D is a single star.
However, since the time that its radial velocity was first measured
\citep{Plaskett1931}, it was considered to be a spectroscopic
binary. Indeed, Vitrichenko (2002b), using all available velocity
data, including his own measurements on IUE spectra, derived
preliminary orbital elements, with two possible solutions for the
orbital period that differ from each other by a factor of two. The
value for the systemic velocity of this object given in Table 5 is
that for the shorter period solution, preferred by the author, but
the error we associate with it is three times larger than that
given in his paper, since it is representative of the difference
between the systemic velocity as derived from both solutions. In
addition, this star is probably a visual binary \citep{Close2012}.

The values for the systemic velocity for $\theta^1$ Ori E and F
come, respectively, from \citet{Costero2008} and Costero et al.
(in preparation). 

\begin{table}[!h]
\caption{Radial velocities of Trapezium stars}
\begin{center}
\begin{tabular}{cc}
\hline
Name & $v$\\
&(km s\textsuperscript{-1})\\
\hline
$\theta^1$ Ori A & $28 \pm 1$\\
$\theta^1$ Ori B & $26 \pm 1$\\
$\theta^1$ Ori C & $23.6 \pm 2.0$ \\
$\theta^1$ Ori D & $32.4 \pm 3.0$\\
$\theta^1$ Ori E & $34.3 \pm 1.0 $ \\
$\theta^1$ Ori F & $20 \pm 4$\\
\hline
\end{tabular}
\end{center}
\label{tvr}
\end{table}%

We stress the fact that the uncertainties in Table~5 are probably gross underestimates. The radial velocities quoted are also very uncertain.
 
As stated above, several of the OT components are multiple stars.
In principle, the orbital motion of a secondary component around
its primary may produce displacements of the primary that could be
confused with relative motions with respect to other components of
the OT. However, we note that the only component whose orbital
motion could be of the order of the DA errors in the measures is
component C. The maximum relative orbital displacement that
component C could show (see Table 3 in \citet{Kraus2009}) is 43
mas over one orbital period. Since the mass ratio of the components is at most 0.23
\citep{Kraus2009} the displacement due to orbital motion of
component C would be at most 10 mas, smaller than our measurement
errors, which are of the order of 20 mas for the relative
position.  All other components have either spectroscopic
companions, or else very large separations with correspondingly
long periods (much larger than the span of our measurements).

Particularly, in the case of component A, the system A1-A2 is
probably a physical binary \citep{Close2012}. Assuming a circular
orbit for the system and $0^\circ$ inclination, a period of about
400 years is deduced from the change in PA of $0^\circ.92 \pm
0^\circ.07$ per year \citep{Close2012}. In this case, taking a
separation of $0.20\arcsec$ and a mass ratio of 0.2
\citep{Weigelt1999}, component A1 will show an oscillation, about
the barycenter, of $0.08\arcsec$ in 200 years. Such an oscillation
is much smaller than the precision of the WDS data, and hence
undetectable.

On the other hand, the WFPC2 measurements are essentially
coeval with those of \citet{Close2012} who report a separation
rate of $-1.4 \pm 0.22$ mas yr\textsuperscript{-1} between A1 and A2; this translates
into a $0.28$ mas yr\textsuperscript{-1} movement of A1 towards the barycenter, which
is an order of magnitude smaller than our reported $2.8 \pm 0.6$
mas yr\textsuperscript{-1} rate of separation between A and E.

The effect of the orbital motion of A1-A2 on the PA between A
and E on the WFPC2 measurements depends on the configuration
between A1, A2 and E which, during the observation period, was
such that the three stars were nearly aligned; consequently, the
PA yearly rate of change reported by \citet{Close2012} for A1-A2
($0^\circ.92$ yr\textsuperscript{-1}) translates into almost $0^\circ.1$ change in 
PA for pair AE during the 12 years lapse of the WFPC2 measurements used. In
nearly 200 years, the change in the PA of the pair of objects A1-E
would oscillate, assuming a circular apparent orbit for A1-A2,
with an amplitude of $0^\circ.9$. These effects are of the same
order as the errors of a single WFPC2 and WDS datum, respectively,
and consequently undistinguishable, as can be noted in Fig. 4.

We conclude that the orbital motions due to multiplicity of the OT members do not 
affect either the measurements carried out by the DA technique, or those of the WDS compilation.

\subsection{Status and internal velocity dispersion of the Orion Trapezium}

The most important kinematical results are as follows:

\begin{itemize}

\item The value we derive for the rate of change of the separation of component E
relative to A, $5.5 \pm 1.2$ km s\textsuperscript{-1}, although
smaller than that reported by \citet{Allen2004} (7.1 km s$^{-1}$
after correcting it for a distance of 414  pc to the ONC), and
\citet{Sanchez2008} (6.9 $\pm$ 1 km s$^{-1}$) is still large
enough for this component to be escaping from the Trapezium.
Indeed, the escape velocity of the OT system is $\sim6$ km
s\textsuperscript{-1} according to \citet{Allen1974b}. The space
velocity of component E relative to A (the vector sum of $5.5 \pm 1.2$ km
s\textsuperscript{-1} for the separation velocity, and $6.3 \pm 2.0$ 
km s\textsuperscript{-1} for the relative radial velocity, see Table~5) is at least 
$8.3 \pm 2.3$ km s\textsuperscript{-1},
which is larger than the escape velocity of the OT.

\item Although component F also appears to be escaping from the system
we were not able to confirm its physical membership to the OT. It
is probably a foreground star, physically unrelated to the
Trapezium. Using the photometric data of \citet{DaRio2009} and \citet{Muench2002} 
we find that the visual total extinction ($A_{v}$) of star F is at least 0.4 mag 
smaller than that of the bright Trapezium components, adopting the spectral 
type of component F (B8) given by \citet{Herbig1950}. 
This result places F in front of the OT, probably  in the foreground veil \citep{ODell2001}, and is consistent with its 
belonging to the abundant stellar population in front of the Trapezium Cluster 
proposed by \citet{Alves2012}.

\item The velocity dispersion we find for components A, B, C
and D is $1.1$ km s\textsuperscript{-1} in one coordinate. 
Component E is escaping, component F is probably a
foreground star, hence, they have not been considered in this
estimate. Of course, a velocity dispersion based on only four
stars is of doubtful significance, but it is consistent with the
OT being a bound system. Even if we include component E, 
the velocity dispersion in one coordinate increases only to 1.8 km s\textsuperscript{-1}.
The value we estimate is
comparable to that found by \citet{vanAltena1988} (1.37 km
s\textsuperscript{-1}, when adjusted to a distance of 414 pc) for
a much larger sample of member stars of the ONC.  This velocity
dispersion is significantly smaller than that found by
\citet{Jones1988} for a larger sample of ONC stars but, as these
authors remark, if they restrict their sample to the brighter
stars, those with magnitudes comparable to the
\citet{vanAltena1988} sample, the conflict practically disappears.
For these stars, \citet{Jones1988} obtain a velocity dispersion of
1.31 km s\textsuperscript{-1} (adjusted to a distance of 414 pc).
The velocity dispersion we find, although from a very small number
of (bright) stars, is comparable to that of \citet{vanAltena1988}
and to the \citet{Jones1988} bright sample. The space motions,
particularly that of component C, remains a controversial subject.
A comprehensive discussion can be found in \citet{ODell2009}.
Briefly, \citet{vanAltena1988} find a motion of $4.8 \pm 0.5$ km
s\textsuperscript{-1} towards PA 142 degrees, significantly larger
than the velocity dispersion of the ONC; \citet{Tan2004}
interpreted this velocity as evidence that the BN infrared and radio source object (which is moving
in the opposite direction) originated in the OT. However, later
studies of the radio sources in the embedded  BN-KL cluster region
\citep{Rodriguez2005,Gomez2005} found three radio sources moving
away from a common point where they coincided some 500 years ago,
and showed that the source used by \citet{Tan2004} as reference
was one of the moving objects. 

Photographic determination of astrometric positions of stars is notoriously difficult and van Altena's tangential motion of 4.8 km s\textsuperscript{-1} is almost certainly too high \citep{ODell2009}. \citet{vanAltena1988} find a poor agreement of their proper motion
of component C with that found by \citet{McNamara1976}. Indeed the
motions obtained by McNamara are, in units of mas yr\textsuperscript{-1}, $\mu_{x} =
-8.3$, $\mu_{y} = -5.7$, quite incompatible with those found by
\citet{vanAltena1988}, $\mu_{x} = +1.4 \pm 0.17$, $\mu_{y} = -1.8
\pm 0.17$. The data from \citet{vanAltena1988} indicate that
components B and C are moving in opposite directions at about 3.5
mas yr\textsuperscript{-1}, a result incompatible with the relative motions of these
stars found by \citet{Allen1974b} who over a time interval of over
125 years, find no significant relative change in their
separation. The results of the present study imply a separation
velocity of component B relative to C of $0.71$ mas yr\textsuperscript{-1}($1.4 \pm 0.9$ km
s\textsuperscript{-1} see Figure 5 and Table 4), also
incompatible with the \citet{vanAltena1988} result.

\item The rest of the components of the OT, present small and random
velocities that resemble the behavior of a bound and virialized
system as found by \citet{Allen1974a,Allen1974b,Allen2004}.

\end{itemize}

\section{CONCLUSIONS}
\label{sec:conclusions}

Diffracto-Astrometry is a novel technique that permits precise
measurements of the relative positions of stars on saturated
images. In this work we have made use of this technique to measure
44 WFPC2 images of the OT over a time period of 12 years
(1995--2007). We found the relative positions of the six
brightest components (A to F) of the OT and were also able to
measure their relative displacements over this time period. To
supplement the astrometric results from the HST data, we used a
compilation of historical data ($\sim$ 200 yrs) for the motion of
these components from the WDS.  We calculated the relative separation
velocities of all the OT components with respect to all the other
components as well as the relative rate of change of their PAs. 
However, the rate of change of the historical PA measurements
 turned out to be unreliable, 
and that determined from DA only is not sufficiently accurate due to the 
short time base.

Our results
yield the kinematics of the OT with a mean precision of $\sim1.0
$ km s\textsuperscript{-1}.
Component E is moving with a velocity larger than the escape
velocity of the system; thus, we conclude that it is in the
process of escaping the OT, as was found by \citet{Allen2004,
Costero2008, Sanchez2008}. The transverse velocity values of the
other components (A, B, C and D) are small and random, mimicking
the behavior of a bound and virialized system as indicated by
\citet{Allen1974a,Allen1974b,Allen2004}. The fact that one
component of the OT is escaping from this system supports the
\citet{Allen1974a} studies of the dynamics of trapezium-type
systems, in which they study the evolution of 30 systems and found
that, in a time period of a million years, 13 of them, almost $50$
\%, would eject one or two components.

\acknowledgments

We are grateful to DGAPA--UNAM for financial support under project
PAPIIT number IN109809 and CONACyT grants I--52081 and I--102284.
This work is based on observations made with the NASA/ESA Hubble
Space Telescope, obtained from the data archive at the Space
Telescope Science Institute. STScI is operated by the Association
of Universities for Research in Astronomy, Inc. under NASA
contract NAS 5-26555. It also has made use of the Washington
Double Star Catalog maintained at the U.S. Naval Observatory.

We acknowledge an anonymous referee for his/her numerous
suggestions, which greatly improved this work.

We wish to express our special gratitude to C.R. O'Dell for fruitful discussions and encouragement.

\end{document}